\documentclass[acmsmall,screen] %review,anonymous]
{acmart}
\usepackage{enumitem}
\usepackage{subcaption}

\usepackage{makecell} % in preamble if not already included
\usepackage{multirow} 
\usepackage[most]{tcolorbox}
\usepackage{pifont}
\usepackage{listings}
\usepackage{xcolor}
\usepackage{adjustbox}
\usepackage{float}
\usepackage{caption}
\usepackage{listings}
\makeatletter
\newcommand{\linelabel}[1]{%
  \begingroup
  \refstepcounter{lstnumber}% advance so \label grabs this line number
  \label{#1}%
  \addtocounter{lstnumber}{-1}% restore the visible counter
  \endgroup
}
\makeatother

\lstdefinestyle{JavaCustom}{
  language=Java,
  linewidth=\linewidth,
  basicstyle=\footnotesize\ttfamily, % was \scriptsize
  keywordstyle=\color{blue}\bfseries,
  stringstyle=\color{brown},
  commentstyle=\color{gray}\itshape,
  numbers=left,
  numberstyle=\tiny\color{gray},
  stepnumber=1,
  numbersep=3pt,
  xleftmargin=0pt,
  framexleftmargin=0pt,
  showspaces=false,
  showstringspaces=false,
  showtabs=false,
  tabsize=2,
  breaklines=true,
  breakatwhitespace=true,
  columns=fullflexible,
  captionpos=b,
  escapeinside={(*@}{@*)},
  aboveskip=2pt, belowskip=2pt, lineskip=-1pt % tighter vertical spacing
}

\title{A Dataset of Reproducible Flaky-Test Failures}

\author{Suzzana Rafi}
\authornote{Both authors contributed equally to this research.}
\affiliation{
  \institution{George Mason University}
  \country{USA}
}

\author{Mahbub-Ul-Hoque Sumon}
\authornotemark[1]
\affiliation{
  \institution{Bangladesh Election Commission}
  \country{Bangladesh}
}

\author{Md Erfan}
\affiliation{
  \institution{University of Alabama}
  \country{USA}
}

\author{Maruf Morshed Khan}
\affiliation{
  \institution{Ministry of Finance}
  \country{Bangladesh}
}

\author{August Shi}
\affiliation{
  \institution{The University of Texas at Austin}
  \country{USA}
}

\author{Wing Lam}
\affiliation{
  \institution{George Mason University}
  \country{USA}
}

\date{May 2025}
\usepackage{balance}
\usepackage{listings}
\usepackage{xspace}
\usepackage{graphicx}
\usepackage[table,svgnames]{xcolor}
\usepackage{booktabs}
\usepackage{url}
\usepackage{framed}
\usepackage[T1]{fontenc}
\usepackage{amsmath} 
\usepackage{xcolor, soul}

\usepackage{calc}
\usepackage{siunitx}

\definecolor{Green}{rgb}{0.6,1,0.8}
\usepackage{tikz}

\newcommand{\Comment}[1]{}
\newcommand{\Space}[1]{}
\newcommand{\Fix}[1]{\textcolor{red}{[[#1]]}}

\newcommand{\suzzana}[1]{}
 \newcommand{\sumon}[1]{}
\newcommand{\Num}[1]{#1} % mark hardcoded numbers that may need to be checked

\newcommand{\MyPara}[1]{\vspace{1pt}\noindent\textbf{#1}.}

\lstdefinelanguage{XML}{
	basicstyle=\ttfamily\footnotesize,
	sensitive=true,
	numbers=left,
	numberstyle=\tiny,
	stepnumber=1,
	numbersep=5pt,
	showspaces=false,
	showstringspaces=false,
	showtabs=false,
	tabsize=2,
	breaklines=true,
	breakatwhitespace=false,
	morestring=[s]{"}{"},
	stringstyle=\color{darkgreen},
	keywordstyle=\color{red},
	morekeywords={xmlns, encoding, type, name},
	comment=[l]{\#},
	commentstyle=\color{gray}\ttfamily
}
\lstdefinelanguage{YAML}{
	basicstyle=\ttfamily\footnotesize,
	sensitive=true,
	numbers=left,
	numberstyle=\tiny,
	stepnumber=1,
	numbersep=5pt,
	showspaces=false,
	showstringspaces=false,
	showtabs=false,
	tabsize=2,
	breaklines=true,
	breakatwhitespace=false,
	morestring=[s]{"}{"},
	keywordstyle=\color{red},
	morekeywords={true,false,null,y,n},
	comment=[l]{\#},
	commentstyle=\color{gray}\ttfamily
}
\newcommand{\CodeIn}[1]{{\texttt{#1}}}
\definecolor{javapurple}{rgb}{0.5,0,0.35} % strings
\definecolor{linenumbergray}{rgb}{0.5,0.5,0.5}
\lstdefinestyle{Java-github}{
        basicstyle=\ttfamily\footnotesize,
        language=Java,
        commentstyle=\color{linenumbergray},
        stringstyle=\color{javapurple},
        keywordstyle=\color{red},
        morekeywords={@Test},
        morecomment=[s][\color{linenumbergray}]{/**}{*/FlakyCat: Predicting Flaky Tests Categories using Few-Shot Learning},
        numbers=left,
        numberstyle=\tiny\color{linenumbergray},
        numbersep=2.5pt,
        moredelim=**[is][\color{javapurple}]{@h@}{@h@},
        morecomment=[f][{\btHL[fill=gitdel]}]-,
        morecomment=[f][{\btHL[fill=gitadd]}]+,
        breaklines=true,
        breakatwhitespace=false,
        breakindent=8pt,
}

\lstdefinestyle{Python-github}{
        basicstyle=\ttfamily\footnotesize,
        language=Python,
        commentstyle=\color{linenumbergray},
        stringstyle=\color{javapurple},
        keywordstyle=\color{red},
        morekeywords={lamda, fail},
        numbers=left,
        numberstyle=\tiny\color{linenumbergray},
        numbersep=2.5pt,
        moredelim=**[is][\color{javapurple}]{@h@}{@h@},
        moredelim=**[is][\btHL]{@i@}{@i@},
        breaklines=true,
        breakatwhitespace=false,
        breakindent=8pt,
}

\newcommand{\Def}[2]{\expandafter\newcommand\csname rmk-#1\endcsname{#2}}
\newcommand{\Use}[1]{\csname rmk-#1\endcsname}

\newcommand{\totalJiraIssue}{10,808}

\newcommand{\toolname}{ReproFlake\Space{/FlakeFlock}\xspace}

\newcommand{\odtestfull}{Order-dependent flaky test}
\newcommand{\odtest}{OD test}

\newcommand{\relevanttest}{OD-relevant test}   %% Need some term
\newcommand{\jiraD}{Jira}
\newcommand{\idoft}{iDoFT}

\newcommand{\idflakies}{iDFlakies}

\newcommand{\RQOne}{RQ1}
\newcommand{\RQOneFull}{What are the challenges associated with reproducing and fixing of flaky-test failures? }

\newcommand{\RQTwo}{RQ2}
\newcommand{\RQTwoFull}{What are the characteristics of the reproducible and fixed flaky tests in our dataset?}
\newcommand{\RQTwoOne}{RQ2.1}
\newcommand{\RQTwoOneFull}{What categories of flakiness\Space{ (e.g., TD, ID, OD)} are the reproducible and fixed tests?}
\newcommand{\RQTwoTwo}{RQ2.2}
\newcommand{\RQTwoTwoFull}{What is the size and location of the code changes between flaky and fixed versions for different types of flaky tests?\Space{, and between flaky and modified flaky versions?}}
\newcommand{\RQTwoThree}{RQ2.3}
\newcommand{\RQTwoThreeFull}{What types of exceptions are most commonly observed from different types of flaky tests?}
\newcommand{\RQThree}{RQ3}
\newcommand{\RQThreeFull}{What is the process to collect other information, e.g., test code coverage,\Space{ and supporting future flaky test analysis and tool development} with our dataset?}

\newcommand{\cmark}{\ding{51}}%
\newcommand{\xmark}{\ding{55}}%

\newcommand{\fixedPatch}{Fixed.patch}
\newcommand{\moreFlakyPatch}{FlakyCodeChange.patch}
\newcommand{\moreFlakyFixedPatch}{FixedCodeChange.patch}

\newcommand{\implDepFull}{Implementation-Dependent (ID)}
\newcommand{\nonIdempOutFull}{Non-Idempotent-Outcome (NIO)}

\newcommand{\orderDepFull}{Order-Dependent (OD)}

\newcommand{\nonDeterministicFull}{Non-Deterministic (NOD)}
\newcommand{\unclassifiedFull}{Unclassified}
\newcommand{\timingFull}{Timing-Dependent (TD)}

\newcommand{\totalidoftOD}{122}

\newcommand{\implDep}{ID}
\newcommand{\nonIdempOut}{NIO}

\newcommand{\orderDep}{OD}

\newcommand{\unclassified}{Unclassified}
\newcommand{\timing}{TD}

\newcommand{\unknown}{Unknown reason}

\newcommand{\typereproduced}{ID, TD, OD, NIO}
\newcommand{\totalIssueNumer}{300}
\newcommand{\issueFlakyTestGeneratedNumer}{63}
\newcommand{\issueFlakyTestGeneratedODNumer}{4}
\newcommand{\issueFlakyTestGeneratedTDNumer}{36}
\newcommand{\issueFlakyTestGeneratedIDNumer}{14}

\newcommand{\issueFlakyTestGeneratedUNCLNumer}{9}

\newcommand{\artifactName}{zip archive}

\newcommand{\totalIdoftTest}{1745}

\newcommand{\totalIdoftIDTestGenerated}{805}

\newcommand{\totalIdoftNIOTestReproducible}{125}

\newcommand{\totalIdoftTestGenerated}{1052}

\newcommand{\totalTestNumber}{1115}

\newcommand{\totalZipNumber}{300}

\newcommand{\totalJiraCoverage}{36}
\newcommand{\totalIdoftCoverage}{590}

\newcommand{\ExampleProjectTD}{apache commons-collections}

\newcommand{\notjavamaven}{Non-Java/Maven project}
\newcommand{\notjavamavencount}{66}
\newcommand{\notflakyUnit}{Not a flaky unit test}
\newcommand{\notflakyUnitCount}{158}
\newcommand{\compilationFailure}{Compilation failure}
\newcommand{\compilationFailureCount}{25}
\newcommand{\notreproducible}{Not reliably reproducible}
\newcommand{\notreproducibleCount}{10}
\newcommand{\excludedCount}{259}
\newcommand{\nonMerged}{Commit SHA not found}
\newcommand{\nonMergedCount}{81}

\newcommand{\compilationFailureCountIdoft}{191}
\newcommand{\notreproducibleCountIdoft}{421}
\newcommand{\excludedCountIdoft}{693}

\newcommand{\totalPercentTD}{3.22}
\newcommand{\totalPercentOD}{11.30}
\newcommand{\totalPercentID}{73.45}
\newcommand{\totalPercentNIO}{11.21}
\newcommand{\totalPercentUnClassified}{0.81} 

\begin{document}

% \settopmatter{printieeeref=false} % Removes citation information below abstract
% \settopmatter{printacmref=true} % Removes citation information below abstract
% \renewcommand\footnotetextcopyrightpermission[1]{} % removes footnote with conference information in first column
% \pagestyle{plain} % removes running headers

\begin{abstract}

Flaky tests pass and fail non-deterministically when run on the same version of code.
Although many techniques have been proposed to detect, debug, and repair flaky tests, reproducing their failures remains a major challenge due to their inherent nondeterminism. 
Many datasets related to flaky tests exist to help researchers study them, but these datasets are often composed of disjoint sets of flaky tests, where each dataset provides some unique information over the others, such as flaky tests of many different categories, failure logs of flaky tests, or flaky tests reported by developers vs. flaky tests found by automated tools.
In this work, we aim to create a reproducible dataset of flaky tests, which are curated from both developer issue reports and a popular dataset of flaky tests. 

Compared to prior flaky test datasets, our dataset is the first to provide (1)~a reproducible environment to compile the flaky tests, (2)~scripts to run the tests to reproduce the failure, (3)~scripts to automatically apply the flaky test fixes and ensure that the test is no longer flaky, and (4)~execution logs of the flaky test passing and failing. 
We present \toolname{}, a dataset of \totalTestNumber{} reproducible flaky tests, spread across four different flaky test categories.
We create guidelines to help others contribute to this reproducible dataset, as well as demonstrate how to use our dataset to understand the challenges with reproducing flaky test failures (i.e., challenges that researchers may face when using any of the prior flaky test datasets), the characteristics (e.g., location of the fix and its correlation with the flaky test category), as well as the difficulties researchers may face in using our dataset to collect additional information (e.g., code coverage) about flaky tests.
Our findings show that error information helps identify flaky test categories and guide repairs, that unresolved compilation failures highlight challenges in building legacy projects, and that knowing typical fix locations helps prioritize repair efforts.

\end{abstract}

\maketitle

\section{Introduction}

%% Regression testing suffers from flaky tests
Developers rely on regression testing, the practice of rerunning tests after every code change, to ensure their code changes do not break existing functionality.
One major problem in regression testing is the presence of \emph{flaky tests}~\cite{Zhang2014JWMLEN,Luo2014HEM}, which are tests that can both pass and fail when run on the same version of code.
When a test suite has flaky tests and a test fails, developers are often misled to believe that the failure indicates a bug introduced in their recent code changes, but this test could have flakily failed regardless of any changes.
Many software development organizations have reported flaky tests as one of their biggest problems~\cite{Kowalczyk2020NGSLM, MediumExpedia, EngFitBitFlakyTests, ChromiumFlakiness, EclipseConContinuousIntegrationGoogle, Memon2017GNDNSM, Ziftci2017R, BlogsGradleFlakyTests, Jiang2017LYX, BlogsTestingInACloud, LamETAL2019ISSTA, Lam2020MST, DeveloperTestVerification, EckETAL2019FSE, Rahman2018R, YouTubeNetflixAutomationTalk, EngineeringSalesforceFlakyTests, WikiSaucelabsFlakyTests, SudarshanThoughtWorks}.

%% There are flaky test datasets
Prior work on developing techniques to mitigate flaky tests also resulted in various flaky test datasets~\cite{AlshammariETAL2021ICSE,AkilETAL2023AST,Lam2019OSMX,BellETAL2018ICSE,dataset:idoft,Luo2014HEM,EckETAL2019FSE,BarbosaETAL2023TSE,Habchi2022HSFPCT}.
Although many datasets related to flaky tests exist, they differ in the types of information they provide(e.g., flaky tests of many different categories~\cite{AkilETAL2023AST,Lam2019OSMX,dataset:idoft}, failure logs of flaky tests~\cite{AlshammariETAL2021ICSE}, or flaky tests reported by developers~\cite{Luo2014HEM,EckETAL2019FSE,AkilETAL2023AST} vs. flaky tests found by automated tools~\cite{AlshammariETAL2021ICSE,Lam2019OSMX,BellETAL2018ICSE,dataset:idoft}).
% one of the largest dataset of flaky tests~\cite{Lam2019OSMX,dataset:idoft})
% various datasets contain useful information that others do not have, e.g., 
% they are limited in the following ways:
% (1)~
% The first study on flaky tests~\cite{LuoETAL2014FSE} studied flaky-test commits 
For example, one of the most popular flaky test datasets is the International Dataset of Flaky Tests (\idoft{})~\cite{dataset:idoft,LamETAL2019ICST}.
\idoft{} includes names of flaky tests detected using automated tools, along with the project the tests are from, the version in which the tests were detected, the category of the flaky test based on the tool that detected the flaky test, and information related to whether the test has been fixed or not.
Since its creation, there has been much work that uses this dataset to evaluate new approaches~\cite{Fatima2023GB,Fatima2024HB,LamETAL2020ICST,LamETAL2020ISSTA,ShantoFlakeSync2024ICSE,rahman2025rankf,Li2023KLS,Li2022ZWS}.
% \Fix{Are there other datasets worth mentioning?}
However, \idoft{} does not have flaky tests reported by developers, nor the failure logs associated with a flaky failure, which other datasets have~\cite{AlshammariETAL2021ICSE, Luo2014HEM,EckETAL2019FSE, AkilETAL2023AST}.
In contrast, \idoft{} also has pull request links to fixes, which other datasets may not have.

%% Prior datasets are limited in an important way
% \Fix{Need some ``twist'' that states how there are weaknesses in prior datasets that we need to address}

%% We propose a new dataset
We propose \toolname{}, a new dataset of reproducible flaky tests obtained from developer issues reporting flaky tests as well as flaky tests collected in prior research (\idoft{}) but with enriched information.
While \idoft{} provides categorized flaky tests and links to pull requests, it lacks compilable and reproducible environments, failure artifacts (e.g., test failure logs and Maven Surefire test reports), and issue reports.
\toolname{} is the first dataset of flaky tests focused on providing an environment to reproduce the flaky test failures.
% , tests that are inherently difficult to reproduce because of their nondeterministic behavior. 
% \Fix{Honestly, a very bold statement, that no other dataset has reproducible flaky tests (especially given something like NonDex that can run to reproduce flaky tests); also, if the tests are not completely reproducible always, is that going to hurt?}
For each flaky test in this dataset, we provide
(1)~an environment (including Java and Maven versions along with project dependencies) to reliably compile the code and test, along with an automated script to run the test and reproduce the failure,  
(2)~the code changes accepted by developers that fix the flaky test,
(3)~a developer-written issue report describing the flaky test (when available), and
(4)~execution logs of the flaky test passing and failing using our scripts and environment.
Each flaky test in \toolname{} is confirmed to be flaky with automated tools and therefore includes failure and fix information.
Our dataset also includes four categories of flaky tests, which encompasses most of the categories in prior datasets, and flaky tests that are reported by developers along with those found only by automated tools. 
We construct \toolname{} by combining flaky tests from two sources: developer-reported issues in \jiraD{} and previously collected flaky tests from \idoft{}. We manually inspect Jira issue reports to identify true flaky tests and retain only Java, Maven-based unit tests. For both \jiraD{} and \idoft{} tests, we require clearly identifiable flaky and fixing commit SHAs and discard tests that fail to compile. We then reproduce flaky failures using automated scripts and package successfully reproduced tests into Docker environments.

\toolname{} enables researchers to tackle many open challenges related to flaky test reproduction, analysis, and repair. 
By providing flaky and fixed versions of flaky tests, scripts that reliably reproduce flaky failures, and a reproducible environment to compile and rerun flaky tests, we allow researchers to better study flaky test behavior. 
Similar to how datasets such as Defects4J~\cite{Just2014JE} and Bugswarm~\cite{Tomassi2019DWBLDVR} enable various automatic program repair techniques that leverages bug reports~\cite{ifixr,getafix,prophet} and the reproducibility of test failures in the datasets, \toolname{} can enable researchers working on automatic repair of flaky tests to learn fix patterns from buggy/fixed code pairs and bug reports.
Furthermore, the failure-related information offers a ground truth for log analysis of flaky execution or classification.
Researchers building tools for flaky test fault localization, categorization, or automated test repair can use our artifacts to train, test, and evaluate their ideas on real-world flaky tests, all in an environment that enables flaky failure reproducibility. 
\toolname{} enhances the current state-of-the-art flaky test datasets with bug reports, flaky and fixed code, Docker containers, failing/passing logs and test execution reports, and pull request details.
Table~\ref{tab:dataset:info} highlights the key differences between existing flaky test datasets and \toolname{}.

\begin{table*}[t]
    \centering
    \caption{Key differences between our \toolname{} dataset and other flaky test datasets.
    \cmark{} and \xmark{} denote that the dataset does and does not, respectively, have such information.}
    \resizebox{1\textwidth}{!}{%
    \begin{tabular}{l|l|c|c|c|c|c|c}
        \textbf{Dataset} 
        & \textbf{Categories} 
        & \makecell{\textbf{Flaky}\\\textbf{Version}} 
        & \makecell{\textbf{Fixed}\\\textbf{Version}} 
        & \makecell{\textbf{Reproducible}\\\textbf{Environment}} 
        % & \makecell{\textbf{Dependencies}\\\textbf{to Compile}} 
        & \makecell{\textbf{Error Logs \&}\\\textbf{Surefire Reports}} 
        & \makecell{\textbf{Pull}\\\textbf{Request}} 
        & \makecell{\textbf{Issue}\\\textbf{Report}} 
       \\ \hline

        iDoFT~\cite{dataset:idoft} 
        & \makecell[l]{ID, OD, NIO, NOD, TD, TZD} 
    & \cmark
    & \cmark
    & \xmark
    & \xmark
    & \cmark
    & \xmark
 \\\hline
       FlakeFlagger~\cite{AlshammariETAL2021ICSE}
        & Does not classify
        & \cmark
        & \xmark
        & \xmark
        & \cmark
        & \xmark
        & \xmark
        \\\hline

         FlakyFix~\cite{Fatima2024HB}
        & ID
        & \cmark
        & \xmark
        & \xmark
        & \xmark
        & \xmark
        & \xmark
        \\\hline

       DeFlaker~\cite{BellETAL2018ICSE} 
       & Does not classify 
       & \cmark
        & \xmark
        & \xmark
        & \xmark
        & \xmark
        & \xmark
     \\\hline
       FlakyCat~\cite{AkilETAL2023AST} 
       & TD, IO, ID, Randomness, Network 
       & \cmark 
       & \xmark 
       & \xmark 
       & \xmark 
       & \xmark 
       & \xmark 
       \\\hline
       Keila et al.~\cite{BarbosaETAL2023TSE} & \makecell[l]{TD, OD, Platform, Randomness,\\
             Resources, ID, Network}& \xmark
        & \xmark
        & \xmark
        & \xmark
        & \xmark
        & \cmark
       \\ \hline
       \toolname{}
    & TD, OD, ID, NIO
        & \cmark
        & \cmark
        & \cmark
        & \cmark
        & \cmark
        & \cmark
\\
    \end{tabular}
    }
    \label{tab:dataset:info}
\end{table*}

Beyond constructing \toolname{}, we also use it to understand (1)~what are the challenges with reproducing flaky test failures (i.e., challenges that researchers may face when using any of the prior flaky test datasets), (2)~what are the characteristics (e.g., location of the fix and its correlation with the flaky test category of flaky tests in \toolname{}, and (3)~how difficult it is to use the dataset to collect additional information (e.g., code coverage) about flaky tests.
Our findings are primarily relevant to researchers and developers working on flaky test reproduction and repair. For example, by analyzing compilation failures and fix locations, our results inform them how to set up reproducible environments more efficiently and where repair efforts should be prioritized (e.g., focusing on test code over code under test for certain categories of flaky tests).
We also study the exception type of the flaky test failures and find that the majority of exceptions in \toolname{} are assertion errors, which we further categorize based on the exception message.
We find that categorization of exception messages can help obtain the flaky test categories, which can be helpful for debugging~\cite{AkilETAL2023AST} and automated fixing~\cite{flakydoctor} of flaky tests.

This paper makes the following main contributions:
\begin{itemize}
    \item We use publicly available issue reports and the existing \idoft{} dataset to create \toolname{}, a new reproducible dataset containing \totalTestNumber{} flaky tests spread across four different flaky test categories of flaky tests.
    \item \toolname{} consists of a Docker environment to compile legacy flaky test projects and run our provided scripts to reproduce flaky failures for each flaky test.
    \item We provide a set of systematic guidelines for how we and others should determine whether a given flaky test is reproducible or not from issue reports and pull request history.
    \item We present an empirical evaluation of the challenges in constructing a reproducible dataset of flaky tests and study various characteristics of flaky tests, providing insights on how future flaky test research can better aid developer's debugging and fixing of flaky tests.
Our artifacts are publicly available~\cite{reproflake2025}. 
\end{itemize}

\section{Background}
\label{sec:background}

Software testing is key to checking the correctness of software.
When a developer runs tests on their system and observe failures, these failing tests should signal to developers that they have a fault in their code that needs to be fixed.
However, often, some tests fail sporadically.
Such tests, that can pass or fail when run on the same version of code, are \emph{flaky tests}~\cite{EckETAL2019FSE, LuoETAL2014FSE}

Understanding and reproducing flaky tests requires both a clear categorization of their causes and familiarity with the tools that can detect, analyze, or reproduce them.
In this section, we first introduce the main categories of flaky tests and then present the tools we use in our study, spanning a range of detection and reproduction techniques.

% Each category requires different tools and strategies for detection and reproduction. In the following subsection, we describe the tools employed in our dataset construction and evaluation process.

\MyPara{Flaky test categories} Flaky test failures may be of different categories, and prior work has defined several categories observed from open-source and industry code.

\begin{itemize}
    \item \textbf{Order-Dependent (OD):} These tests' pass/fail outcomes depend on the order in which they are run~\cite{Lam2019OSMX, Zhang2014JWMLEN}.
Other tests, known as \textbf{\relevanttest{}s (ODR)}~\cite{rahman2025rankf, Shi2019LOXM}, changes some shared state (e.g., static variables) between the ODR and OD test, thereby influencing the pass/fail outcome of the OD test.
    \item \textbf{\timingFull:} These tests can fail due to concurrency or waiting on asynchronous computations~\cite{ShantoFlakeRake2024ICST, ShantoFlakeSync2024ICSE}.
Uncontrolled thread interleavings due to differences in timing per execution can result in different test outcomes.
    \item \textbf{\implDepFull:} These tests assume deterministic outcomes from APIs whose specifications are actually nondeterministic, such as iteration order over data structures like \CodeIn{HashMap}, whose order is not guaranteed~\cite{Shi2016GLM}.
    \item \textbf{\nonIdempOutFull:} These tests fail when they are run in the same JVM more than once; the test passes when run the first time in the JVM but fails when run again~\cite{WeiETAL2022ICSE}.
These failures arise because the test leaves behind or modifies shared state (heap, file system, etc.), so running it again without resetting that state leads to a different (failing) result.
\end{itemize}

\MyPara{Flaky Test Reproduction and Analysis Tools} We rely on several flaky test detection and reproduction tools to reproduce the failures of different categories of flaky tests.

\begin{itemize}
\item \textbf{iDFlakies~\cite{Lam2019OSMX} for \odtest{}s:}
To detect and reproduce \odtest{}s, we use iDFlakies, a customized Maven plugin for running tests in different orders.
iDFlakies runs tests in random orders, where a test that passes in one order but fails in another is flaky.
We can easily reproduce a flaky test failure by rerunning a failing order.
Given a failing and passing order, prior work~\cite{Shi2019LOXM} relied on delta-debugging~\cite{Zeller1999} to systematically search for the ODR test corresponding to the \odtest{}.

\item \textbf{NonDex~\cite{Shi2016GLM, GyoriETAL2016FSEDemo} for \implDep{} tests:}
NonDex identifies APIs with nondeterministic specifications, such as those with uncontrolled iteration order, and changes their output to something different but valid, such as randomizing the iteration order.
If a test fails when run with a different order of its output, then it is flaky.

\item \textbf{iDFlakies~\cite{Lam2019OSMX} for \nonIdempOut{} tests:}
\idflakies{} detects \nonIdempOut{} tests by running the same test multiple times in the same JVM using a custom test runner; if the test passes on the first execution but fails on a subsequent run, it is classified as an \nonIdempOut{} flaky test.

\item \textbf{Docker:} To provide a reliable environment for reproducing flaky test failures, we utilize Docker~\cite{Docker} images.
We can configure a Docker image to contain the relevant operating system as well as source code and dependencies needed for execution.
We can then share this Docker image with others, who can use it to reproduce specific executions.
\end{itemize}

\section{Example Flaky Test}

\begin{figure*}[t]
\centering
\begin{subfigure}{0.48\textwidth}
\begin{lstlisting}[style=JavaCustom, basicstyle=\tiny\ttfamily]
@Test public void testSave() throws IOException { (*@\linelabel{line:comments}@*)
  final String comments = "Hello world!";  {(*@\linelabel{line:actual}@*)
  try (ByteArrayOutputStream actual = new ByteArrayOutputStream()) {(*@\linelabel{line:savecoll}@*)
    PropertiesFactory.EMPTY_PROPERTIES.save(actual, comments);
    try (ByteArrayOutputStream expected = new ByteArrayOutputStream()) {
      PropertiesFactory.INSTANCE.createProperties()
        .save(expected, comments);
      String expectedComment = getFirstLine(expected.toString("UTF-8"));
      String actualComment   = getFirstLine(actual.toString("UTF-8"));
      assertEquals(expectedComment, actualComment, () ->
        String.format("Expected String '%s' with length '%s'",
          expectedComment, expectedComment.length()));
      expected.reset();
  // (*@\textcolor{red}{Added to simulate flakiness and ensure the test fails}@*)(*@\linelabel{line:start}@*)
  // (*@\textcolor{red}{try \{ Thread.sleep(2000); \} catch (InterruptedException e) \{ \}}@*)(*@\linelabel{line:end}@*)
      try (PrintStream out = new PrintStream(expected)) { (*@\linelabel{line:saveJdk}@*)
        new Properties().save(out, comments);(*@\linelabel{line:asserteroor}@*)
        (*@\textbf{\texttt{assertArrayEquals(expected.toByteArray(), actual.toByteArray(), expected::toString);}}@*)
    } catch (UnsupportedEncodingException e) { fail(e.getMessage(), e); }
  }
}\end{lstlisting}
\caption{Example of \timing{} flaky test}
\label{fig:exampleTD}
\end{subfigure}
\hfill
\begin{subfigure}{0.49\textwidth}
\begin{lstlisting}[style=JavaCustom, basicstyle=\tiny\ttfamily]
@Test public void testSave() throws IOException {
  final String comments = "Hello world!";
  try (ByteArrayOutputStream actual = new ByteArrayOutputStream(); ByteArrayOutputStream expected = new ByteArrayOutputStream()) {
    PropertiesFactory.EMPTY_PROPERTIES.store(actual, comments);
    PropertiesFactory.INSTANCE.createProperties().store(expected, comments);
    String expectedComment = getFirstLine(expected.toString("UTF-8"));
    String actualComment = getFirstLine(actual.toString("UTF-8"));
    assertEquals(expectedComment, actualComment, () ->
      String.format("Expected String '%s' with length '%s'", expectedComment, expectedComment.length()));
    expected.reset();
// (*@\textcolor{red}{Added to simulate flakiness in fixed version, but test should still pass}@*)(*@\linelabel{line:threadstart}@*)
// (*@\textcolor{red}{try \{ Thread.sleep(2000); \} catch (InterruptedException e) \{ \}}@*)(*@\linelabel{line:threadend}@*)
    try (PrintStream out = new PrintStream(expected)) {
      new Properties().store(out, comments);
    }
    String[] expectedLines = expected.toString(
      StandardCharsets.UTF_8.displayName()).split("\\n");
    String[] actualLines = actual.toString(
      StandardCharsets.UTF_8.displayName()).split("\\n");(*@\linelabel{line:fixstart}@*)
    (*@\textbf{assertEquals(expectedLines.length, actualLines.length);} @*)(*@\linelabel{line:fixend}@*)
    (*@\textbf{assertEquals(expectedLines[0], actualLines[0]);} @*)
  }
}\end{lstlisting}
\caption{Developer fixed version of the flaky test}
\label{fig:exampleFixTD}
\end{subfigure}
\caption{Example of \timing{} flaky and fixed test from \ExampleProjectTD{}. The red-commented block was added to better reproduce the failure, and it successfully turns the intermittent failure into a deterministic one in the flaky version, while no test failures occur in the fixed version.}
\label{fig:exampleTD-all}
\end{figure*}

We identified a \timingFull{} flaky test for our dataset while examining issue report \emph{COLLECTIONS-812}~\cite{issueReportTd:collection} from the Apache Commons-Collections project.
The flaky test is shown in Figure~\ref{fig:exampleTD}, and the developer’s fixed version is shown in Figure~\ref{fig:exampleFixTD}. 
This test compares the behavior of the \CodeIn{save} method when invoked on an empty, immutable \CodeIn{EmptyProperties} object defined in Apache Commons-Collections versus the standard JDK implementation, \CodeIn{java.util.Properties}.

In Figure~\ref{fig:exampleTD}, the \CodeIn{save} method writes three elements to the provided \CodeIn{OutputStream} in Line~\ref{line:savecoll} and \CodeIn{PrintStream} in Line~\ref{line:saveJdk}:  
(1)~a String \CodeIn{comments} (Line~\ref{line:comments}),
(2)~the current date and time (recorded to the nearest second),
and (3)~key-value entries on subsequent lines (with no entries for \CodeIn{EmptyProperties}). 

Assume that the \CodeIn{OutputStream} object \CodeIn{actual} at Line~\ref{line:actual} is written at some time $T_0$ by the method \CodeIn{EMPTY\_PROPERTIES.save(...)} on Line~\ref{line:savecoll}.
Then, the \CodeIn{OutputStream} object \CodeIn{expected} is written at some other time $T_1$ by the \CodeIn{new Properties().save(...)} call on Line~\ref{line:saveJdk}.
If $T_0$ and $T_1$ happen within the same second, the test passes; if they happen in different seconds due to delays in the test code or the code under test (CUT), the assertion on Line~\ref{line:asserteroor} fails, because it compares the entire \CodeIn{OutputStream} rather than only the first line. 
The pass/fail outcome thus depends solely on the wall-clock timing of the \CodeIn{save} call.

We are unable to reliably reproduce the flaky test failure in every run of the test.
However, if we insert a delay, \CodeIn{Thread.sleep(2000)} at Line~\ref{line:start}, before the call to \CodeIn{new Properties().save(...)}, we can ensure that $T_1 \geq T_0 + 2\,\mathrm{s}$, so the times always differ by at least one second.
As such, the \timing{} flaky test failure now fails 100\% of the time, making the flaky failure now a deterministically reproducible one.

In Figure~\ref{fig:exampleFixTD}, we show how the developer later fixed the flaky test by changing the assertion to check that the two outputs have the same number of lines and that only the first line matches (Lines~\ref{line:fixstart}--~\ref{line:fixend}).
Now, even if we introduce a delay at the same location (the red-commented \CodeIn{Thread.sleep(2000)} on Line~\ref{line:threadstart}), the test no longer fails, regardless of the amount of time between the two points $T_1$ and $T_2$.

We include this flaky test in our dataset, along with all the code and information needed to help a user compile and run the flaky test, for both the flaky version where the test could fail as well as the fixed version where the test should no longer fail.
To help with confirming the flakiness behavior as well as it being fixed, we also include the changes needed to reliably reproduce the failure, namely the inserted \CodeIn{Thread.sleep} delay needed to make this test fail.
In other words, for each flaky test in our dataset, we include four code versions: Flaky, Fixed, FlakyCodeChange (e.g., Flaky with \CodeIn{Thread.sleep} to reproduce failure), and FixedCodeChange (e.g., Fixed with \CodeIn{Thread.sleep} to confirm the failure no longer occurs).

\section{\toolname Infrastructure and Artifact}

We present \toolname~\cite{reproflake2025}, which is a dataset of reproducible flaky tests of multiple categories. 
We curate the dataset with flaky tests from a popular flaky test dataset, \idoft ~\cite{dataset:idoft}, and issue reports from popular open-source projects posted in \jiraD{}. Our dataset consists of \artifactName{}s that provide dockerized environments to reproduce flaky-test failures across multiple categories, including \implDepFull, \orderDepFull, \timingFull, and \nonDeterministicFull tests.

\subsection{Dataset Structure}

For each Maven module\footnote{In Maven, a module is a directory within a project that contains its own pom.xml and is built and tested as a separate unit.} that contains a flaky test, \toolname has a \artifactName{} for it for a particular code version. 
If a module has multiple flaky tests in it, only one zip archive is available for that, which includes the required data to reproduce all flaky tests from that module. 
The zip archive contains the following information to help reproduce the flaky test failures:
\begin{itemize}
    \item Source code of the project at the version when the test/tests in the module were last found to be flaky so that we can work on the flaky version to reproduce the failure. 
    
    \item The Maven dependency repository (i.e., the \CodeIn{.m2} directory) which contains all dependencies needed to compile and run the test for different code versions. We kept it in a single folder named \CodeIn{Flakym2} to reduce the data size, as there is often little to no difference in the dependencies among the different code versions due to their similarity.
    
    \item The patch file, \CodeIn{Fixed.patch}, which a user can apply to the flaky version of the project to get the developer fixed version. The developer fixed version contains the fixed flaky test.
        
    \item The patch file, \CodeIn{FlakyCodeChange.patch}, which a user can apply to the existing flaky version, changing it into a version with higher rate of failure. This patch is only relevant for \timing{} flaky tests as other categories have deterministic failure reproduction scripts.
    
    \item The patch file, \CodeIn{FixedCodeChange.patch}, which a user can apply to the fixed version of the code to see if it still fails or if the failure rate increases. (Only applicable for TD flaky tests.)

    \item An information file containing metadata for the flaky test, including the issue report link, the GitHub commit SHA, and the pull request link.

\end{itemize}

We provide a framework to automatically compile and run the test, reproducing the flaky failure deterministically. 
Our framework will use our \artifactName{} and run the test in dockerized environment to generate a ``results'' folder containing the result of the test being run in different code versions, where each version has the test run logs, Maven Surefire reports, and a summary file that summarizes the results (number of failure/pass) of multiple test runs, as well as the coverage information for the test run.

\begin{figure}[t]
    \centering
    \includegraphics[width=0.90\columnwidth]{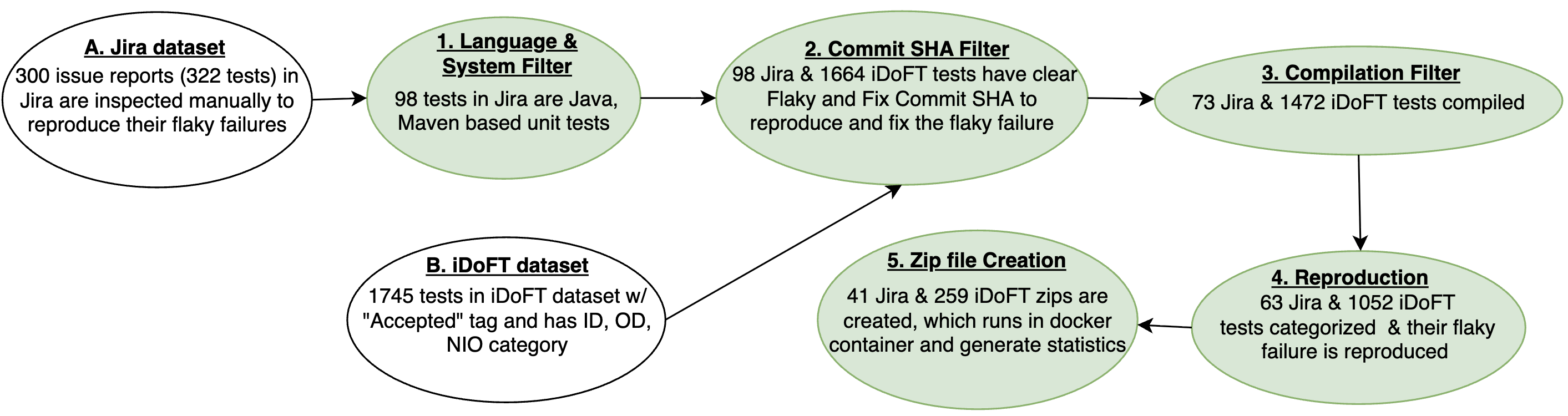}
    \caption{High-level workflow of our methodology}
    \vspace{-4ex}
    \label{fig:workflow}
\end{figure}

Our work involves mining of two data sources: a)~\jiraD {} Issue Report~\cite{dataset:jira} and b)~\idoft{} dataset~\cite{dataset:idoft}.
\jiraD{} is a widely used issue-tracking platform, especially for Apache projects, whereas \idoft{} (International Dataset of Flaky Tests) is a popular dataset that provides a large and diverse collection of real-world flaky tests. In Figure~\ref{fig:workflow}, we can see that A and B represent the two datasets. We curate our reproducible dataset from these sources by following the steps outlined in the workflow diagram.
The process starts with collecting data from \jiraD{} issue reports and \idoft{} dataset, followed by filtering out tests based on language, relevant information availability, compilation, reproducibility, and zip creation process.

\subsubsection{\jiraD {} Issue Report}

\paragraph{\textbf{A. \jiraD{} Dataset (Input).}}
 To obtain flaky-test-related issue reports from developers, we mined \jiraD{} for issue reports that explicitly mention flaky behavior using the keywords ``intermittent'', ``flaky'', ``flak'' - either in the title, description, or comments.
% These reports form a key source for constructing our flaky test dataset.
We used the \jiraD{} Python library~\cite{jira_python} to query issue reports from all Apache projects on \jiraD{}.
%
% \paragraph{Data Inclusion Criteria}
% We collected a set of issues containing the keywords ``flaky'', ``flaky test'', ``flak'', ``intermittent'' from Jira issues. 
Our search resulted in \Num\totalJiraIssue{} issues.
The complete list of issues we obtained from \jiraD{} is available online in our artifact~\cite{reproflake2025}. Among the issues we obtained from \jiraD{}, we randomly selected around 3\% of the issue reports (\totalIssueNumer{}) and distributed them among four reviewers for manual inspection.

\paragraph{\textbf{Inspection Methodology \& Dataset Creation Guideline}}

To create a reliably reproducible dataset of flaky tests, we created and followed a comprehensive issue inspection guideline.
All reviewers independently followed the guideline, and updated the guideline as needed (in which case all reviewers were notified and updated their labeling accordingly). 
We make our guideline available online~\cite{guideline25} so future work on flaky tests can follow it to create additional reproducible flaky test artifacts.

To ensure consistency and reliable reproducibility, we followed a five-step process, represented in Figure~\ref{fig:workflow} and described in detail in the guideline. The workflow bubbles follow the guideline.
In some cases, a single bubble may consist of multiple subsections of the guideline, but the section number always corresponds to the bubble number. For example, Language and System Filter in bubble 1 maps to Section 1.1 and 1.2 in our guideline.

Bubble 1 in the workflow diagram checks whether an issue involves a Java–Maven flaky unit test. Bubble 2 identifies the relevant commits and test information (i.e., module, fully qualified test name), Bubble 3 handles compilation, Bubble 4 classifies and reproduces flaky tests, and Bubble 5 creates and validates the final data artifacts.

\paragraph{\textbf{1. Language \& System Filter.}} 

We first verified whether each issue corresponds to a Java, Maven-based project and a flaky unit test. Reviewers inspected the project repository to confirm the presence of a pom.xml file and Java source code, and analyzed the issue report, comments, and pull request details to determine whether the issue involves flaky unit-test behavior rather than integration or system-level flakiness. Issues failing this check were excluded from further analysis.

\paragraph{\textbf{2. Commit SHA Filter.}}

For valid Java, Maven-based flaky unit tests, we identified the developer-fixed merged commit, and its immediate preceding commit, which we treated as the flaky version. To locate these commits, reviewers first searched for the issue ID in the project’s GitHub repository, examining associated pull requests and commits. When a merged pull request was available, the merge commit was treated as the fixed version and its immediate predecessor as the flaky version.

\paragraph{\textbf{3. Compilation Filter.}}

We attempted to compile each module corresponding to the flaky/fix commit using Maven. Compilation issues were mostly resolved by running Maven with multiple skip flags to avoid unnecessary checks, switching between Java 8, 11, and 17 to resolve version issues, replacing SNAPSHOT dependencies with the closest stable versions from Maven Central in all pom.xml files, and updating repository URLs from http to https to ensure secure downloads. Issues that could not be compiled after these steps were excluded.

\paragraph{\textbf{4. Reproduction.}}

The main differences in reproducing various categories of flaky tests lies in the distinct methodologies used for their reproduction. More details about the flaky test category and tools to detect them are given in Section~\ref{sec:background} of the paper. Below, we summarize the approaches used to reproduce each category of flaky test we came across while mining the \jiraD{} issue reports.
%\paragraph{}
\par
\textbf{\timingFull{} flaky tests}: \timing{} flaky tests are those that exhibit non-deterministic behavior due to variations in execution timing, such as thread scheduling, race conditions, or asynchronous operations. We tried to simulate the exact behavior that makes the test flaky by inserting a sleep either in the test code or the code under test. The major challenge here is where to add \CodeIn{Thread.sleep(delay)}. We performed manual analysis of the issue report along with its comment section, the pull request and commit in GitHub, the changed code to resolve the flakiness, the test method, code under test, and other associated code with this test, and then tried adding delay manually in different portions of the code and observed the test execution. Based on this analysis, we manually inserted Thread.sleep() near the code where timing or synchronization affects the test result, such as before assertions or around asynchronous operations. 

If adding the delay caused a significant increase in the failure rate out of 10 runs compared to runs without the delay, we assumed the test to be a \timing{} flaky test. We then increased the delay by 100 ms in each iteration until the test failed in all 10 runs, ensuring deterministic reproduction of flakiness in our environment. If increasing the delay did not make the test fail deterministically, we did not label it as \timing{} flaky test and considered it under other flakiness categories.

\par \textbf{\odtestfull{}:}
\odtestfull{}s can be reproduced in two ways:

\begin{itemize}
    \item \textbf{Unknown OD-relevant test} If we do not know the OD-relevant test, we may use tools like \idflakies{}, which runs the tests in a test suite in random orders to expose flaky failures.
    Since \idflakies{} relies on a customized test runner, which showed some incompatibilities with a few JUnit tests in our dataset, we instead made small changes to the Maven Surefire plugin, customizing it to allow running the tests in specific orders.
    We first gather all tests in a module and execute them in 20 randomized orders using the custom Maven Surefire plugin.
    If a test shows nondeterministic results (at least one pass and one fail across runs), we then rerun it in both a failing and passing order five times each.
    If the test consistently fails in the failing order and passes in the passing order, we confirm it as an OD flaky test.  
    \item \textbf{Known OD-relevant test:} If the \relevanttest{} is known from the report or pull request, we reproduce the flakiness by running the \odtest{} and its \relevanttest{} in the specific execution order required for the \odtest{} to fail, using our customized Maven Surefire plugin.
\end{itemize}

\par\textbf{Implementation-Dependent (ID) Tests:} ID tests are reproduced by exposing nondeterministic behavior in APIs or execution environments with tools such as NonDex~\cite{tool:nondex}, which explores different allowed behaviors of under-determined Java APIs (like \CodeIn{HashMap} iteration order) and reports tests that fail under these variations.
For the failure run, we store the NonDex seed that caused the failure, which represents the specific randomization configuration that caused the failure, allowing the same behavior to be reproduced deterministically in later runs. We use this stored seed during the reproduction step to guarantee deterministic reproduction.

If the category-specific tool fails, we run all other tools; if none succeed, we execute the test 100 times (Section 4.3 of the guideline). Tests that fail intermittently are labeled Unreliably Reproducible (with a presumed category when available, or Unclassified otherwise), while tests that never fail are marked Unreproducible.

\paragraph{\textbf{5. Zip File Creation.}}

To build zip archives for the different categories of flaky tests in Section 5 (bubble 5) of the guideline, we first prepare the project by cloning the repository and creating directories for the flaky and fixed versions.
Then we construct patches to capture differences across versions, including \fixedPatch{} for the fixed version, \moreFlakyPatch{} for the more-flaky version for TD flaky tests only, and \moreFlakyFixedPatch{} for the developer-fix to check if the flakiness persists. 
We generate patch files by computing the source-code differences between the \texttt{Flaky} and \texttt{Fixed} versions.
We also cache the local Maven dependencies for all code versions in a single directory called Flakym2, along with the information about the flaky test and patches with code changes. The test is then executed multiple times to collect statistics. We package the final setup into a ZIP archive and add it to our dataset for easy reproduction.

We add the generated \artifactName{} that contains all the relevant information for our infrastructure by updating a central CSV configuration file.
The CSV records key data for each flaky test, including its name, category, project module, corresponding ZIP and folder names, its \relevanttest{} (for order-dependent cases), the number of iterations run, test configuration, and Java version used.
Our infrastructure then uses the CSV and ZIP file to extract the project, place it in an appropriate Docker image for the test category, and run the tests (using tools/multiple reruns/code changes) on different code versions to gather results.

To ensure consistency across inspection, we maintain a structured CSV log of all relevant information.
Each row corresponds to a flaky test and includes information about the reproduction steps such as the issue and commit details, compilation status, presumed flaky category from the report, tools/code changes used, results from multiple runs (including 100-run statistics), logs and failure rates, final detected category, and supporting artifacts like patch files, NonDex seeds, test orders, and reproducibility status.

\subsubsection{\idoft{} Dataset}

\paragraph{\textbf{B. \idoft{} Dataset (Input).}}We also reproduced flaky tests from the International Dataset of Flaky Tests (\idoft{})~\cite{dataset:idoft}, because it provides a large and diverse collection of real-world flaky tests, representing various categories of flakiness observed and fixed in well-known, Java, Maven-based open-source projects. The methodology for reproducing flaky tests from iDoFT is illustrated in Figure~\ref{fig:workflow}, it starts from bubble B. 
Unlike the Jira dataset, the iDoFT dataset 
(i) already provides the flaky-test category through the \textit{category} column, 
(ii) exclusively contains flaky unit tests from Java, Maven-based projects, allowing us to skip the language and system filtering step (bubble~1), 
(iii) includes only tests associated with accepted pull requests, and 
(iv) contains \nonIdempOutFull{} flaky tests. 

We filtered \totalIdoftTest{} flaky tests with \textit{Accepted } status {where the pull requests containing the fixes were accepted}. After working with these flaky test we found \totalIdoftTestGenerated{} flaky tests are reliably reproducible of categories \implDep, \nonIdempOut, \orderDep, and \unclassified{} (UD/NOD), and \excludedCountIdoft{} flaky tests are excluded due to \compilationFailure{}, \nonMerged{}, and  \notreproducible{}.

\paragraph{\textbf{Inspection Methodology}}

%A combination of manual inspection and automation was applied to verify the correctness of flaky classifications. Scripts were used to confirm test class names, method names, and reproduce failures across different commits using the associated PRs.
For the iDoFT dataset, the commit SHA filtering step differed slightly. 
As the \idoft{}  dataset already contained accepted pull request link, we directly obtained the merged commit from the pull request as the fixed version and used its preceding commit as the flaky version, without requiring issue-based commit searches. 
Issue reports are not explicitly included in \idoft{}. Therefore, we traced back to its corresponding issue report by automatically matching issue IDs in pull request descriptions, followed by manual inspection when automated mapping was unsuccessful.
Also, for NIO tests, we used the custom testrunner provided by \idflakies{}, configured to execute the same test twice within the same JVM. A test was classified as \nonIdempOut{} if it passed on the first execution but failed on the second execution in the same JVM.
All remaining steps - compilation filter, reproduction, zip file creation - followed the same workflow as for the \jiraD{} dataset.

\subsection{Dataset Statistics}

The dataset comprises \totalZipNumber {} \artifactName {} covering \totalTestNumber{} reproducible flaky unit tests across the categories \typereproduced{} and \unclassified. Each ZIP file may contain multiple flaky tests associated with a specific merged commit SHA and a particular module within the same project. Table~\ref{tab:dataset-outcomes} summarizes the composition of our final dataset, detailing how flaky tests from the \jiraD{} and \idoft{} sources were filtered, reproduced, and included. It reports the number of tests per flaky-test category, along with the reasons for exclusion including compilation errors, missing reproduction metadata (e.g., commit or polluter information), and unreproducible failures. The unreproducible flaky tests mostly come from always passing or failing and not showing flakiness. 

\begin{table}[t]
\centering
\caption{Summary of flaky-test outcomes from \jiraD{} and \idoft{} datasets.}
\label{tab:dataset-outcomes}
\small
\setlength{\tabcolsep}{6pt}
\begin{tabular}{l l r}
\toprule
\textbf{Outcome} & \textbf{Reason / Ctegory} & \textbf{\# Tests} \\
\midrule
\multicolumn{3}{l}{\textbf{Jira Dataset}} \\
\midrule
Included & \timingFull{} & \issueFlakyTestGeneratedTDNumer{} \\
Included & \implDepFull{} & \issueFlakyTestGeneratedIDNumer{} \\
Included & \orderDepFull{} & \issueFlakyTestGeneratedODNumer{} \\
Included & \unclassifiedFull{} & \issueFlakyTestGeneratedUNCLNumer{} \\
\textbf{Included Total} &  & \textbf{\issueFlakyTestGeneratedNumer{}} \\
\midrule
Excluded & \notjavamaven{} & \notjavamavencount{} \\
Excluded & \notflakyUnit{} & \notflakyUnitCount{} \\
Excluded & \compilationFailure{} & \compilationFailureCount{} \\
Excluded & \notreproducible{} & \notreproducibleCount{} \\
\textbf{Excluded Total} &  & \textbf{\excludedCount{}} \\
\midrule
\multicolumn{3}{l}{\textbf{iDoFT Dataset}} \\
\midrule
Included & \implDepFull{} & \totalIdoftIDTestGenerated{} \\
Included & \orderDepFull{} & \totalidoftOD{} \\
Included & \nonIdempOutFull{} & \totalIdoftNIOTestReproducible{} \\
\textbf{Included Total} &  & \textbf{\totalIdoftTestGenerated{}} \\
\midrule
Excluded & \nonMerged{} & \nonMergedCount{}\\
Excluded & \compilationFailure{} & \compilationFailureCountIdoft{} \\
Excluded & \notreproducible{} & \notreproducibleCountIdoft{} \\
\textbf{Excluded Total} &  & \textbf{\excludedCountIdoft{}} \\
\bottomrule
\end{tabular}
\end{table}

\Comment{\section{Dataset Creation}
\label{sec:methodology}

We used the JIRA Python library (which wraps the JIRA REST API) to query a public JIRA instance (Apache’s JIRA) and export a set of issues. We used the keywords “flaky”, “flaky test”, “flak”, “intermittent” to query JIRA issues. We initially had 10808 issues (last queried in June 2024). We decided to assign each person a batch containing up to 20 issues drawn from 15 projects at a time. We distributed 939 issues in total among 5 people.

In some cases, developer assumes that increasing the wait value in test code would resolve the problem. If developer increases the wait time randomly, it will increase the CI/CD time in total. }
\section{Experimental Setup and Evaluation}
\label{sec:setup}

We address the following research questions to showcase the challenges with creating a flaky test dataset, provide insights into what future flaky test research should focus on and how they may do so, and the extensibility of our dataset.

\noindent\textbf{\RQOne{}}: \RQOneFull{}

\noindent \textbf{\RQTwo{}}: \RQTwoFull{}
    \begin{itemize}
        \item \textbf{\RQTwoOne{}}: \RQTwoOneFull{}
        \item \textbf{\RQTwoTwo{}}: \RQTwoTwoFull{}
        \item \textbf{\RQTwoThree{}}: \RQTwoThreeFull{}
    \end{itemize}

\noindent \textbf{\RQThree}: \RQThreeFull{}

RQ1 is important because it highlights the challenges in creating a reproducible dataset, e.g., missing information or compilation issues, and offers guidance in how researchers should avoid such challenges. 
\RQTwo{} explores what the reproducible and fixed flaky tests look like - what kind of flakiness they have, what code changes were made, what types of errors they cause, etc. 
These details will help researchers build better tools for detecting or fixing flaky tests. \RQThree{} looks at how useful our dataset is for collecting code coverage and running further analysis. By making it easier to gather this data, we hope to support future research and tool development around flaky tests. Our dataset opens several directions for future research on flaky tests detection and repair, which are described in later subsections.

Knowing the type of flakiness helps developers understand what causes it, e.g., timing, input, or order issues, and fix the flakiness correctly. 
Comparing code changes between flaky and fixed versions shows where and how much code was changed and whether the fix worked. The dataset offers a ready-to-use setup with Docker, and tools like JaCoCo to run tests, measure coverage, and compare versions. Our RQ results combine findings from both datasets.

\subsection{\textbf{\RQOne{}: Challenges faced while reproducing flaky tests}}
\label{sec:error_classification}

We faced several challenges while trying to reproduce flaky tests. We could overcome some challenges, others we could not.

There were 216 compilation errors that we could not fix, but for some, we were able to resolve them by running Maven with multiple skip flags to avoid unnecessary checks, switching between Java 8, 11, and 17 to resolve version issues, replacing SNAPSHOT dependencies with the closest stable versions from Maven Central in all pom.xml files, and updating repository URLs from http to https to ensure secure downloads. 

Other challenges consist of not always finding the issue reports related to the PRs in \idoft{} dataset. We addressed this by automatically linking pull requests to issue IDs and manually inspecting cases where automated mapping failed. 

A key challenge we overcame was identifying the fixing commit on the main branch corresponding to each pull request. We manually inspected GitHub to locate merged commits when available. Another challenge we overcame was reproducing TD flaky tests. 

Using tools like FlakeRake~\cite{ShantoFlakeRake2024ICST} appeared to be difficult due to insufficient documentation and version compatibility issues. The tool required a specific version of Java to run, and our projects were not always compatible with it. Even when they were compatible, using this tool was very expensive, took a long time to get the exact place to inject delays for one TD test. We overcame these issues by manually investigating the code and finding appropriate places to inject the delays.

Few of the challenges that we could not resolve were in some cases while trying to find the merged commit from a fixed PR, pull requests were closed or the changes were incorporated differently, making it impossible to identify an exact fixing commit. Issue reports could not be linked to some PRs because they were unavailable. Also, the 216 compilation issues we could not solve, but tried to analyze the error categories in the next paragraph to understand them better. We did not apply automated compilation-repair tools (e.g., Maven dependency repair tools), leaving their evaluation as future work. In many cases (431 tests), we could not reproduce the flaky behavior, even after running tools that usually catch these issues. For 157 OD tests, 259 ID tests, and 5 NIO tests, tools that run tests in random order certain times did not help, possibly because the flaky order was not triggered.

\textbf{Compilation Error Categories:}
As the majority of failures are compilation-related, we performed a compilation error categorization on the combined \idoft{} and \jiraD{} datasets. Compilation errors are not always experimental design problems, these errors come from the projects’ own outdated or unreachable libraries and repositories. 
We randomly sampled 100 build logs and manually assigned 15 fine-grained labels, each representing a distinct compilation error type (i.e., subcategory) corresponding to a recurring root cause (e.g., missing dependencies, Java version mismatches). Characteristic error messages and keywords from each label were used to construct regex rules and to build a few-shot prompt for GPT-4o-mini. 

New logs were classified using a regex-first approach; logs that did not match any rule were classified by GPT using the few-shot examples selected from the manually labeled sample. The resulting subcategories were then grouped into higher-level categories based on shared failure causes (e.g., Dependency Resolution Errors, Plugin Execution Errors). Logs that remained unclassified were re-examined, leading to refined rules or new labels, and this process was repeated until most logs were covered and the categories stabilized.

Our \textbf{Dependency Resolution Error} analysis shows what the most dominant category of compilation errors is in both datasets.
Dependency resolution errors account for a big portion of the errors - \textbf{80.0\%} of the compilation errors (20 out of 25) in the \jiraD{} dataset and \textbf{50.8\%} of the compilation errors (98 out of 191) in the \idoft{} dataset. 
These errors occur when Maven is unable to find, resolve, or download required dependencies or artifacts needed during the build. 
Some common issues are missing artifacts, wrong or missing version numbers, problems reaching repositories, or issues caused by missing or broken parent POMs.

Among the dependency resolution errors, \textit{Blocked or non-existent HTTP repositories} encompass \textbf{31.6\%} of the errors, which are blocked or outdated HTTP repositories that prevent artifact downloads.

A close second problem (30.8\%) is \textit{Missing Dependency Version in POM}, where dependencies are declared without explicit version information, leaving the build system unable to resolve the correct artifact.
Other issues include \textit{Missing Dependencies}, where required libraries cannot be found, failed dependency resolution from version mismatches or network issues, dependency management conflicts between library versions, and non-resolvable parent POMs that break inheritance. 

\begin{tcolorbox}[
  colback=gray!5,
  colframe=black,
  boxrule=0.5pt,
  sharp corners,
]
\textbf{Open Research Challenge - \textit{Improving Compilation of Legacy Projects}}

We classify compilation error categories, which can guide future research in understanding and addressing common build issues encountered when compiling legacy projects used for flaky test research.
Our findings show that most issues are dependency resolution errors, which occur when required libraries or repositories cannot be found or accessed, often due to missing dependency version information in POM files, outdated or blocked HTTP repositories, missing dependencies, or broken parent POM links.
These findings suggest that improving dependency availability and updating repository links should be prioritized to resolve these build failures.\end{tcolorbox}

\begin{figure}[t!]
    \centering
   % \begin{minipage}{0.40\textwidth}
  
    %\end{minipage}\hfill
    \begin{minipage}{0.50\textwidth}
        \centering
        \includegraphics[width=\linewidth]{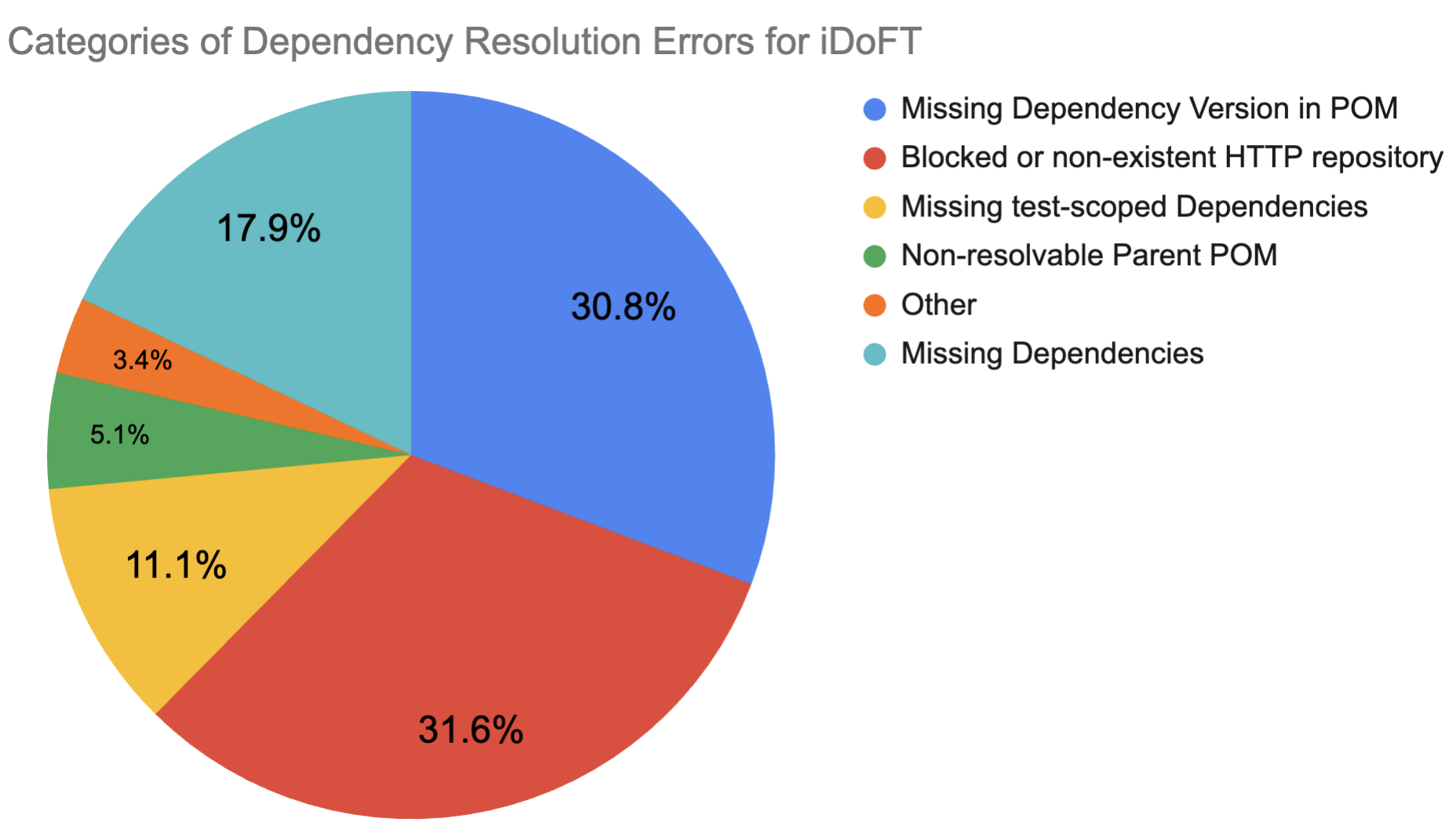}
        %\label{fig:res_error}
    \end{minipage}
 %   \vspace{-2ex}
    \caption{Categories of Dependency Resolution Errors in our Dataset (N=216)}
    \label{fig:res_error}
    \vspace{-2ex}
\end{figure}

%\subsubsection{}
\subsection{{\RQTwo}: Characteristics of Reproducible and Fixed Flaky Tests}
     Our goal is to understand the characteristics of our reproducible flaky test dataset. We aim to explain the categories of flaky tests we could reproduce, the location and size of the changes needed to fix these flaky tests, and categories of flaky exceptions. Our findings aim to provide insights into what future flaky test research should focus on and what such research can do to address reproduction and fixing challenges.
\subsubsection{{\RQTwoOne}: Categories of Flakiness in Reproducible and Fixed Tests}
      
  Our reproducible dataset includes different categories of flaky tests - for \idoft{} dataset, we reproduce \typereproduced{} flaky tests. For the \jiraD{} dataset, we are able to reproduce \implDep{}, \orderDep{}, \timing{}, and a few unclassified ones for \unknown{}. Table~\ref{tab:dataset-outcomes} has the breakdown of the type of flaky tests and the count for each category. The majority of the dataset consists of Implementation-Dependent flaky tests for the \idoft{} dataset, whereas for \jiraD{}, it is mostly \timing{} flaky tests. We reproduce \textbf{TD} (\totalPercentTD\%), \textbf{OD} (\totalPercentOD\%), \textbf{ID} (\totalPercentID\%), \textbf{NIO} (\totalPercentNIO\%), and \textbf{Unclassified} (\totalPercentUnClassified\%) flaky tests, where the percentages are calculated across the entire set of \totalTestNumber{} reproducible tests from both \idoft{} and \jiraD{} datasets.

\subsubsection{{\RQTwoTwo}: Size and Location of Code Changes Between Flaky and Fixed Versions}

We analyze where fixes occur - test code versus code under test - and how large those fixes are across different flakiness categories to identify repair location patterns and guide future automated repair. By seeing patterns across different flaky types, developers can better predict the effort needed and quickly focus on the right part of the code. 

As shown in Table~\ref{tab:avg_change_test_cut}, for OD, TD, UD, and NIO tests, fixes primarily modify the test code, with relatively small changes to the code under test. These fixes typically involve a limited number of lines and are localized to the test code.
For ID tests, however, fixes show more variation. 
On average, ID fixes involve larger changes to the code under test than to the test code, indicating that some ID cases require substantial production code modifications. 
At the same time, the median ID fix changes more test code than production code, indicating that most ID fixes are test-centric, while a smaller number of large production code changes substantially increase the average. 
Nevertheless, our results suggest that ID flakiness can stem from both test-level issues and deeper implementation-level problems.
Changes are measured as summation of additions and deletions of lines. Overall, 93.53\% of ID, 28.57\% of OD, 58.33\% of TD, 99.2\% of NIO, and 66.67\% of Unclassified tests changed only one location - either the test code or the production code. 
Future research on fixing flaky tests can benefit from first focusing on which part of the code the fix is likely to go in before attempting different patches.

\begin{tcolorbox}[
  colback=gray!5,
  colframe=black,
  boxrule=0.5pt,
  sharp corners,
]
\textbf{Open Research Challenge - \textit{Prioritizing Location of Fixes}}

Our findings in \RQTwoTwo{} show that most flaky tests, regardless of category, fixes mostly occur in the test code rather than code under test (Table~\ref{tab:avg_change_test_cut}).
These findings can help developers prioritize the location of fix or guide LLM-based repair systems or automation repair tools to focus their fixes in the most likely location first. For example, if a flaky test is categorized as TD, OD or NIO, the developer/model can start by suggesting fixes in the test code (e.g., waits or synchronization for TD). If it is Implementation-Dependent (ID), they can also inspect the code under test for shared-state or order-sensitive logic as well as the test code. 
\end{tcolorbox}

\begin{table}[h!]
\centering
\caption{Average Change in Test Code and Code under Test across Flaky Types}
\label{tab:avg_change_test_cut}
% \resizebox{\columnwidth}{!}{%
%\begin{tabular}{|l|r|r|}
\begin{tabular}{|l|p{3cm}|p{3cm}|}

\hline
\textbf{Type} & \textbf{Avg Line Changes in Test Code} & \textbf{Avg Line Changes in Code under Test} \\ \hline
ID  & 112.35 & 121.05 \\ \hline
OD  & 11.06 & 1.59 \\ \hline
NIO & 52.18 & 0.13 \\ \hline
TD  & 39.36  & 18.97 \\ \hline
UD  & 46.00 & 11.67 \\ \hline
\end{tabular}%
 % }
\end{table}

\subsubsection{{\RQTwoThree}: Categories of Flaky Exceptions across Flaky Categories}
\label{sec:rq23_flaky_exception_categories}

We categorize the flaky test failures for all types of flaky tests in our dataset. 
Knowing if certain error log types (e.g., Timeout, AssertionError, NullPointerException) consistently lead to certain fix strategies (e.g., adding waits, reordering tests, initializing state) can help guide what automated techniques should do to fix flaky tests using the error log. 
We followed the same approach described in Section~\ref{sec:error_classification} for categorizing compilation errors to classify the flaky failures. 
Namely, we manually labeled 100 failure logs with fine-grained failure subcategories, derived regex rules from recurring log patterns, and used few-shot prompts with GPT-4o-mini to classify logs that did not match any rule. 

The resulting subcategories were then systematically grouped into higher-level categories based on semantic similarity of failure causes (e.g., order mismatches, value mismatches, structural mismatches), and the classification was iteratively refined until the categories stabilized and covered most failure patterns.

\begin{figure}[t]
    \centering
\begin{subfigure}{0.45\textwidth}
  %  \begin{minipage}{0.48\textwidth}
        \centering
        \includegraphics[width=1.07\linewidth]{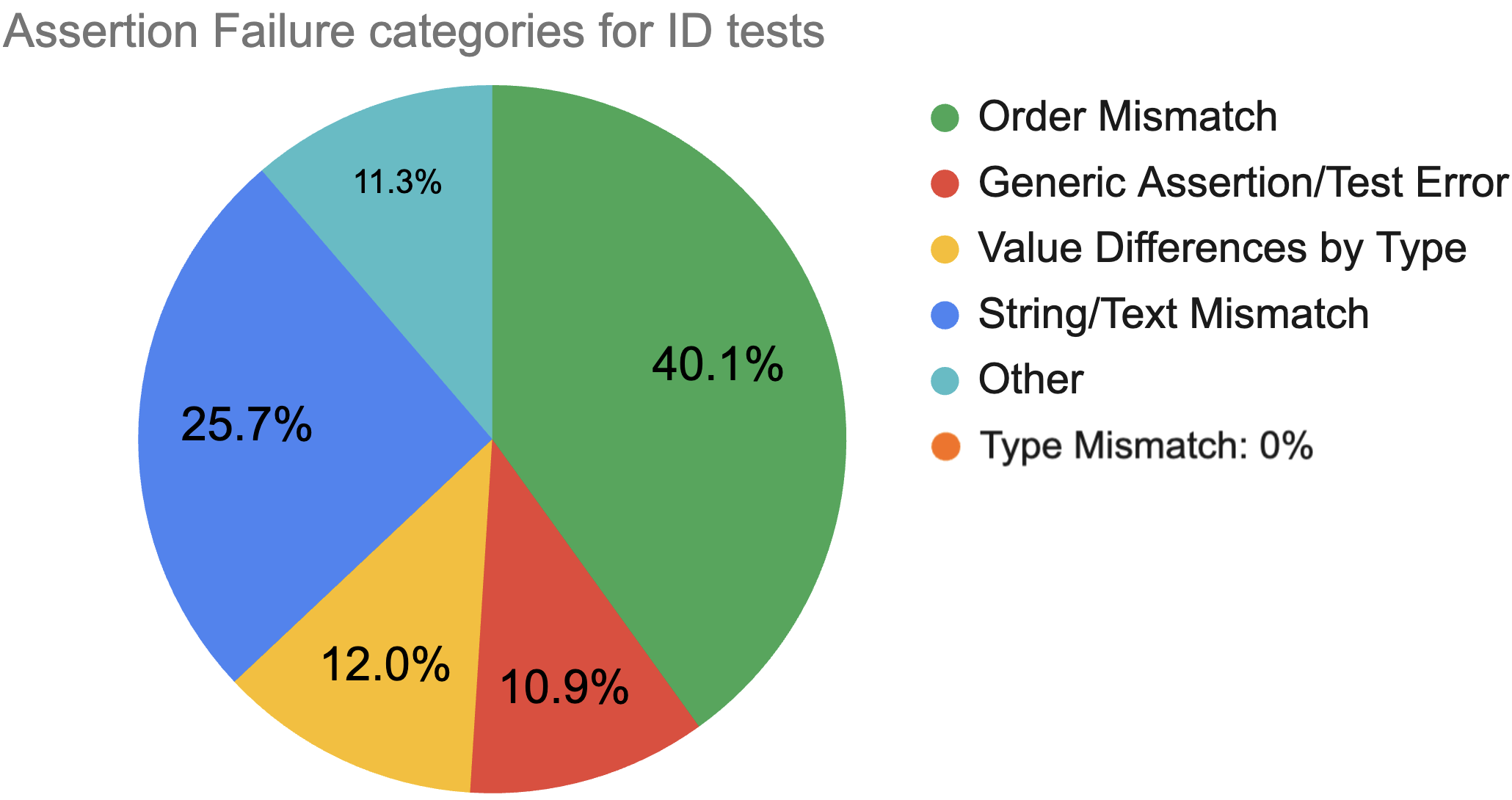}

         \end{subfigure}
\hfill
\begin{subfigure}{0.47\textwidth}
        %\caption{Assertion error categories}
   % \end{minipage}\hfill
    %\begin{minipage}{0.48\textwidth}
        \centering
        \includegraphics[width=0.90\linewidth]{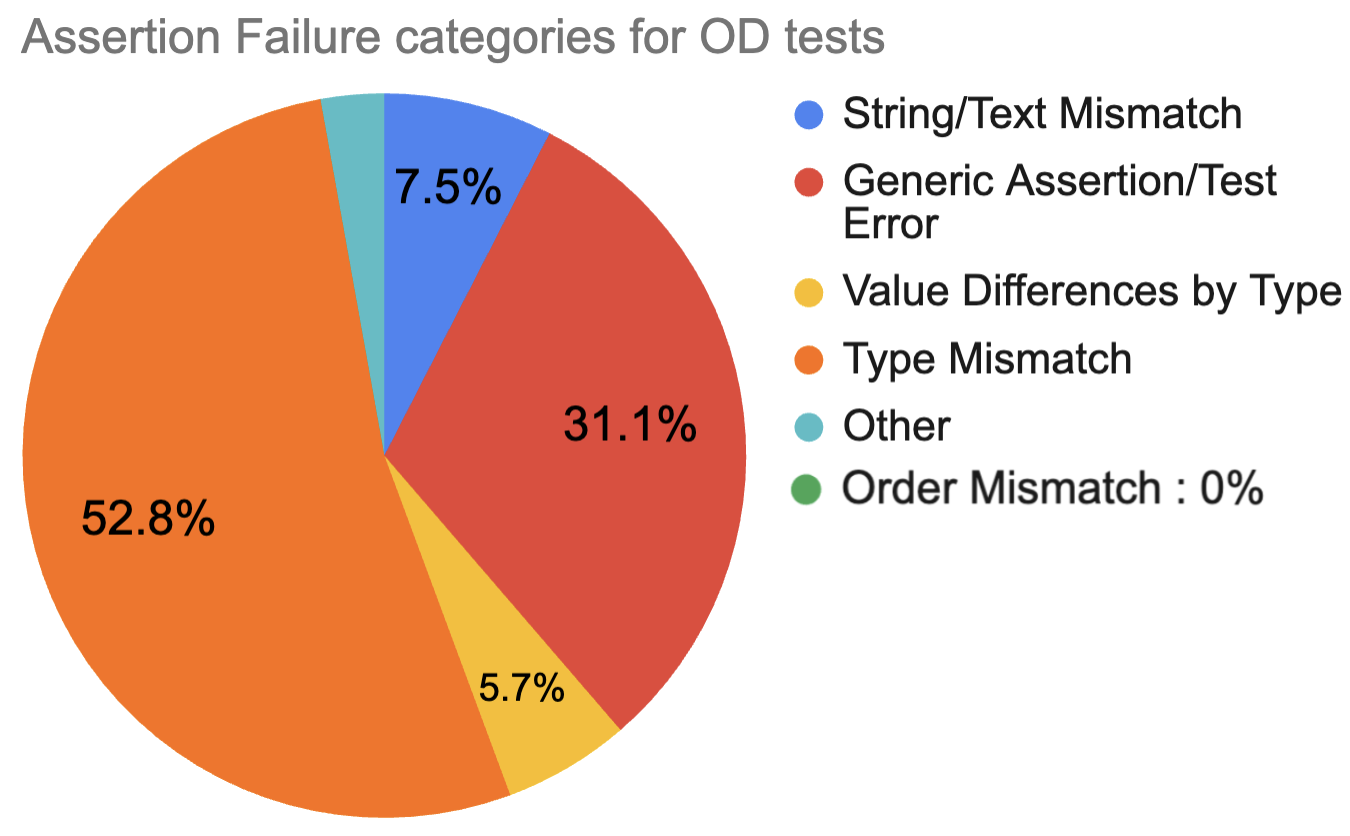}
        % \caption{Assertion error categories for \odtest{} tests in \idoft{}}
 \end{subfigure}
          \caption{Assertion error categories for ID (N=805) and OD (N=122) flaky tests in our dataset.}
        \label{fig:id_od}
    %\end{minipage}
\end{figure}

\begin{figure}[t]
\begin{subfigure}{0.45\textwidth}

    \centering
    \includegraphics[width=\linewidth]{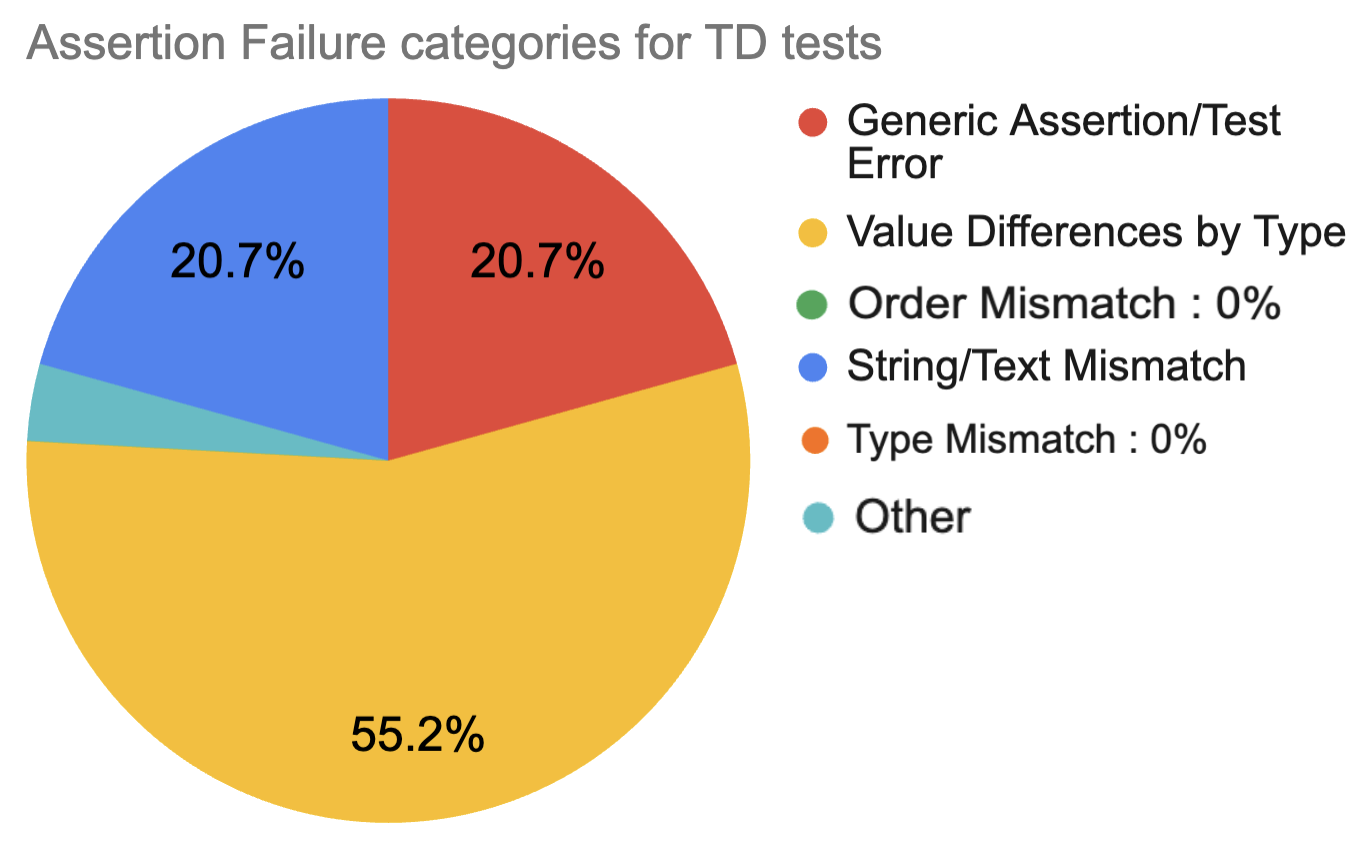}
         \end{subfigure}
\hfill
\begin{subfigure}{0.45\textwidth}
    
    \includegraphics[width=\linewidth]{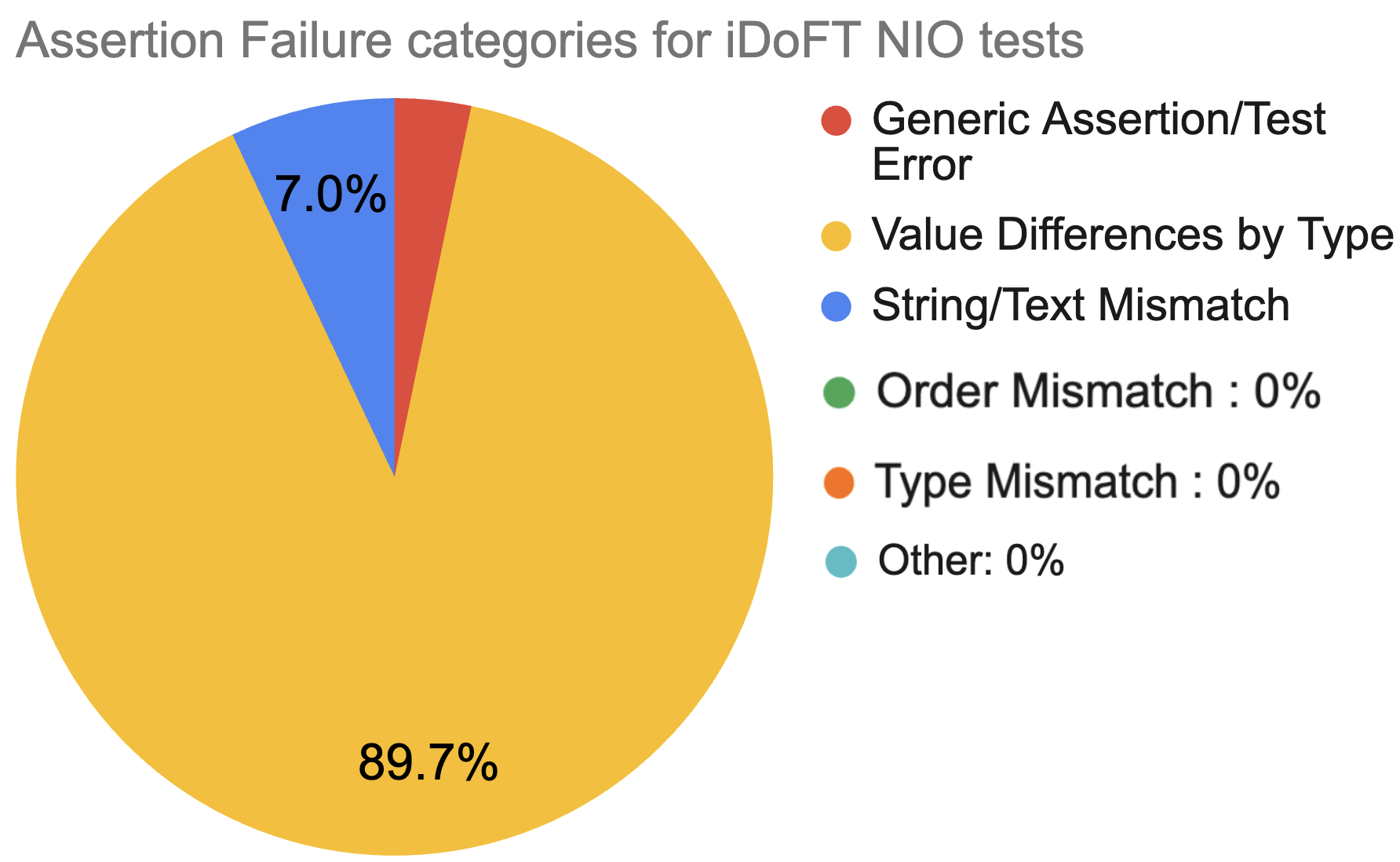}
    \end{subfigure}
    \caption{Assertion error categories for TD (36) and NIO (125) flaky tests in our dataset.}
    \label{fig:td_nio}
\end{figure}

In the \idoft{} dataset, nearly all ID test failures result from \textit{Assertion Errors} (99.4\%). 
OD tests also fail mainly because of \textit{Assertion Errors} (96.3\%), with a few cases involving missing class definitions or RPC errors. A Remote Procedure Call (RPC) error occurs when a test calls a method on an external system or service, but the call fails due to connection or setup issues. 
For NIO tests, \textit{Assertion Errors} are also the primary cause (98.9\%), with some errors due to global state conflicts or registry errors.

In the \jiraD{} dataset, TD tests mostly exhibit \textit{Assertion Errors} (90.6\%), with some \textit{Timeouts} (6.3\%) caused by delays in async operations and a few \textit{NullPointer Exceptions} (3.1\%). 
Both ID and OD tests show only \textit{Assertion Errors}. 
In the \textit{Unclassified} category, 66.7\% failures are \textit{Assertion Errors} and 33.3\% are due to \textit{Project Configuration} problems like broken build files or missing setup elements.

As most flaky errors are Assertion errors across all categories, we have further sub categorized the assertion errors. We can see in Figure~\ref{fig:id_od} and \ref{fig:td_nio}, for \textbf{ID} tests, the most common issue is \textit{order mismatch} (40.1\%). In \textbf{OD} tests, \textit{type mismatch} dominates (52.8\%). For \textbf{TD} tests, \textit{value difference by type} accounts for 55.2\%, while in \textbf{NIO} tests, it is at 89.7\%. 
Understanding these detailed assertion error types can help prioritize debugging and fix strategies.

Some of the prevalent subcategories are described below:

\begin{itemize}
  \item\textbf{Order Mismatch} - The expected and actual elements are correct, but they appear in a different order 
  
  \item \textbf{Type Mismatch} - The expected and actual values are of different types or classes (e.g., different data types, object types, or exception types).
  
  \item \textbf{Value Difference by Type} - The type matches, but the actual value differs (e.g., 58 vs.\ 57)  
  
  \item \textbf{Structure Mismatch} - The overall data structure differs (e.g., list vs.\ set, or tree shape changes)  
  
  \item \textbf{String/Text Mismatch} The expected and actual strings contain the same data but differ in text format or order (e.g., JSON fields appear in a different order)
  
  \item \textbf{Generic Assertion/Test Error} - A broad failure with little detail (e.g., \texttt{assertTrue} failing without context). They can indicate missing context or weak assertions  
\end{itemize}

\textit{Example (ID vs. OD):} Order or formatting differences indicate ID flakiness, while unexpected exceptions or behavior indicate OD flakiness.\\
ID $\rightarrow$ \texttt{expected:<\{"a":1,"b":2\}>, was:<\{"b":2,"a":1\}>}\\
OD $\rightarrow$ \texttt{expected:<RuntimeException>, was:<AssertionError>}

\begin{tcolorbox}[
  colback=gray!5,
  colframe=black,
  boxrule=0.5pt,
  sharp corners,
]
\textbf{Open Research Challenge - \textit{Inferring Flaky Test Types from Error Categories}}

A potential use of \RQTwoThree{} analysis is to take an arbitrary error message and using the observed frequency of flaky test categories in real world projects~\cite{LamETAL2020ICSE, Luo2014HEM}, estimate the likelihood that the failure belongs to a particular flaky test category. If one type of error is dominant in more than one category of flaky tests, we can calculate the likelihood according to the frequency of that flaky test category in practice, and estimate which category the failure is most likely to belong to.
Prior work such as FlakyCat~\cite{AkilETAL2023AST} and FlakyDoctor~\cite{flakydoctor} shows that knowing the flaky test category is useful for diagnosis and repair, but typically requires analyzing test code or observing test behavior across multiple executions. Our results show that for a nontrivial number of flaky tests, the category can be inferred using error log messages alone, which can subsequently be used for fixing flaky tests.\end{tcolorbox}

\begin{tcolorbox}[
  colback=gray!5,
  colframe=black,
  boxrule=0.5pt,
  sharp corners,
]
\textbf{Open Research Challenge - \textit{Providing New Information to Fix Flaky Tests}}

It is possible to combine the information from \RQTwoTwo{} and \RQTwoThree{} and feed an LLM the flaky error message, with error category, suggested fix location for that category, combined with additional context such as covered methods and issue report descriptions to automatically generate a repair. Our work provides issue report information (if available), as well as covered methods and line numbers for flaky and fixed versions. Unlike prior tools such as FlakyFix~\cite{Fatima2024HB}, which predicts fix types from test code, or FlakyDoctor~\cite{flakydoctor}, which requires known OD/ID labels and static localization, users can use the error category to determine category of flaky test and consequently repair the test.

\end{tcolorbox}

\subsection{ \textbf{\RQThree: Process for Extending the Dataset with Additional Tools}} We want to assess the framework's usability by showing how easily we can collect code coverage using the infrastructure. We can run tests, collect code coverage, and compare different versions using the scripts in our framework. 
We added other tools like JaCoCo and assessed its usability. 
To add tools, we provide a script that automatically modifies the POM files of projects (configuration file for Maven-based projects). 

We used JaCoCo to obtain code coverage for \totalJiraCoverage{} tests out of \issueFlakyTestGeneratedNumer{} tests from \jiraD{} issue reports and \totalIdoftCoverage{} tests out of \totalTestNumber{} tests from \idoft{} dataset.

The average line coverage of the flaky version divided by the fixed version by type of flaky test is:
0.992 for ID, 1.0 for OD, 0.992 for TD, 0.997 for NIO, and 1.042 for Unclassified.
Values near 1.0 show that fixes mainly stabilized execution without changing which code ran. Ratios < 1.0 (ID, TD, NIO) mean the fixed version covered slightly more lines, suggesting the fix enabled additional or previously skipped paths. 
Ratios > 1.0 (unclassified) indicate that the flaky version ran extra or redundant code. 
Overall, OD remained close to 1.0, showing that their fixes improved determinism or isolation rather than coverage.
The fact that flaky and fixed versions cover almost the same code means that most problems are not what code runs, but how or when it runs (like timing, shared state, or order issues). 

\begin{tcolorbox}[
  colback=gray!5,
  colframe=black,
  boxrule=0.5pt,
  sharp corners,
]
\textbf{Open Research Challenge - \textit{Comparing Flaky Test Fixes}}

Another research direction is to compare LLM-generated fixes with developer-written patches from our dataset. Future studies can analyze differences in coverage, mutation score, failure rate, and lines of code changed to assess repair quality. Since LLMs can often generate multiple potential fixes, coverage data from our dataset can be used to identify which generated fix best matches real developer fixes. Comparing coverage similarity can help determine which LLM-generated fix is closest to common developer fixes. 
\end{tcolorbox}

\section{Threats to Validity}

Flaky tests can produce nondeterministic results due to multiple categories.
In our dataset, we follow a systematic guideline to identify the correct categories of a flaky test.
The actual categories of a flaky test may still change over time (e.g., fixed flaky test can become flaky again on certain days of the week).
To help mitigate this threat, we ran our reproducible environment multiple times for each flaky test to ensure that the failures are reliably reproducible and provide the error logs of the failure and scripts used to obtain the failure.

The findings of our study may not generalize to other flaky test categories.
Our dataset composed of four main categories of flaky tests, but some studies~\cite{Luo2014HEM,EckETAL2019FSE} have reported 10+ categories of flaky tests.
The categories of flaky tests we included in our dataset and study was dictated by the flaky tests that we can find and reliably reproduce, which was dictated by what categories of flaky tests have automated tools to help with failure reproduction.
Our literature review revealed automated reproduction tools only for the four categories that we included in our study.
To help mitigate this threat, we include a systematic guideline in our artifact so others can more easily replicate our work for other flaky tests of the categories we studied or even for categories that we did not study.

\section{Related Work}
\label{sec:related}

Datasets for improving software engineering research have a long tradition. 
From manually-curated datasets for test adequacy criteria~\cite{Hutchins1994} in 1994 to more recent and general datasets for software testing and analysis, such as SIR~\cite{dataset:SIR}, Defects4J~\cite{dataset:defects4j}, and BugSwarm~\cite{dataset:bugswarm}.\Comment{
Some examples on related topics include 
Defects4J~\Fix{\cite{}}, a dataset of reproducible bugs in code under test, 
BugSwarm~\Fix{\cite{}}, a dataset of bugs where one version of code has failing tests while another version has passing tests
}
There has also been an increasing number of domain specific datasets, such as for test prioritization research\cite{Fallahzadeh2022a},\Space{ AndroZoo for} research related to Android apps~\cite{Allix:2016},\Space{ TravisTorrent for} continuous integration related research~\cite{Beller2017},\Space{ Bug prediction dataset} and for bug prediction research~\cite{DAmb2010a}. 

\toolname, like prior datasets, provides real-world software artifacts for evaluating testing techniques. However, unlike general-purpose datasets such as Defects4J or BugSwarm, it focuses specifically on flaky tests and supports reproducibility by including execution environments, logs, and both flaky and fixed versions. This makes it well-suited for studying nondeterministic test behavior and repair.
\Comment{\Fix{include 2-3 sentences here describing how your dataset is similar and different than these existing datasets}{\Fix{Refer back to the table in intro and make sure you cite each of the rows in that table here}}}
A widely used flaky test dataset is \idoft{}~\cite{dataset:idoft}, which consists of list of flaky tests, their GitHub URL, flaky commit, their pull request information, but lacks reproducibility artifacts such as error logs, issue reports, and executable environments (Table~\ref{tab:dataset:info}). 
FlakeFlagger~\cite{AlshammariETAL2021ICSE} includes a different set of flaky tests than \idoft{} and extends the information by including build and failure logs, but still does not provide fixed versions or reproducible environments (Table~\ref{tab:dataset:info}). Similarly, FlakyFix~\cite{Fatima2024HB} and DeFlaker~\cite{BellETAL2018ICSE} include flaky tests but lack key artifacts such as fixed versions, logs, and reproducibility support. FlakyCat~\cite{AkilETAL2023AST} categorizes flaky tests across multiple dimensions but does not include reproducibility artifacts or complete debugging information. Keila et al.~\cite{BarbosaETAL2023TSE} provide issue reports and categorization, but do not include flaky or fixed versions or reproducible setups.

Compared to all of these datasets, \toolname{} is the only dataset (Table~\ref{tab:dataset:info}) that simultaneously provides flaky and fixed versions, reproducible environments, detailed error logs, pull requests, and issue reports, enabling end-to-end reproducibility and deeper analysis of flaky test behavior.

\section{Conclusion}
\label{sec:conclusion}

In the past decade, many software development organizations have reported flaky tests as one of their biggest problems.
To help with this issue, many research papers have studied the problems of flaky tests and proposed automated techniques to detect, debug, and fix flaky tests.
To facilitate such research, various datasets of flaky tests have been proposed, yet reproducing the failures of flaky tests, especially for many categories of them, can still be challenging.
Different datasets have also curated different types of information, e.g., error logs and issue reports, which are often useful for automated debugging and fixing techniques.
In this paper, we present \toolname{}, which is a dataset of \totalTestNumber{} flaky tests focused on providing the following for each flaky test (1) an environment (in the form of Docker images) to aid in the reproducibility of the flaky test failure, (2) fix that was accepted by developers, (3) error log that details the message and stack trace of the flaky test failure, and (4) pull request and issue report information.
Using this dataset, we studied the challenges in reproducing flaky test failures, the characteristics of flaky tests, and the difficulty in using our dataset to collect additional information.
Our study revealed several important open research challenges, which our dataset can help address:
(A)~compilation of legacy projects can focus on the prominent categories that were responsible for why some subjects in our study did not compile, 
(B)~fixing of flaky tests can leverage the category of a flaky test to better predict where the fix is likely to be, 
(C)~debugging of flaky tests can leverage the error category of a flaky test failure to predict what category a flaky test is, 
(D)~fixing of flaky tests can leverage a vast combination of information altogether (e.g., flaky test category, error log, issue report) from our dataset to improve fixing, and  
(E)~comparison of flaky test fixes using traditional testing metrics (e.g., code coverage, mutation score) can be done easier with our dataset.
We make our dataset publicly available~\cite{reproflake2025} and provide guidelines for transparency in how we constructed the dataset and to enable future work to expand it as needed.

\section{Data Availability}

All relevant data, scripts, and results generated for this work are available on our website~\cite{reproflake2025}. 

\bibliographystyle{ACM-Reference-Format}
\bibliography{ref,wing-bibs/flaky-tests-links,wing-bibs/flaky-tests-papers,wing-bibs/testing,wing-bibs/crossrefs}

\end{document}